\documentclass[journal]{IEEEtran}
\usepackage{cite}
\usepackage{graphicx}
\usepackage{amsmath}
\usepackage{subfigure}
\usepackage{multirow}

\usepackage[normalem]{ulem}
\useunder{\uline}{\ul}{}
\usepackage[flushleft]{threeparttable}
\usepackage{algorithm}
\usepackage{algpseudocode}
 
\makeatletter
\def\BState{\State\hskip-\ALG@thistlm}
\makeatother

\begin{document}

\title{A Minimum Spanning Tree Representation of Anime Similarities}

\author{\IEEEauthorblockA{Canggih Puspo Wibowo \\
		Sagasitas Research Center \\
		Yogyakarta, Indonesia \\
		(canggih.p.w@ieee.org)}}

%\markboth{Working Paper [Version 1]}%
%{Judul}

\maketitle

\begin{abstract}
In this work, a new way to represent Japanese animation (anime) is presented. We applied a minimum spanning tree to show the relation between anime. The distance between anime is calculated through three similarity measurements, namely crew, score histogram, and topic similarities. Finally, the centralities are also computed to reveal the most significance anime. The result shows that the minimum spanning tree can be used to determine the similarity anime. Furthermore, by using centralities calculation, we found some anime that are significant to others.

\end{abstract}

\begin{IEEEkeywords}
anime, similarity measurement, minimum spanning tree
\end{IEEEkeywords}

\IEEEpeerreviewmaketitle

\section{Introduction}

Minimum spanning tree is an undirected graph that has no cycles, connects to every vertex, and has the minimal total weighting for its edges. It is known as a graph which has low complexity and easy to implement \cite{Herman2000}. Mostly, minimum spanning tree is used to represent wires, roads, and water pipes so that the total cost is minimum. However, recently it has been used in various areas such as geographical information \cite{Moncla2016}, radio networks \cite{Murmu2015}, EEG \cite{Crobe2016}, chip architecture \cite{Maican2015} and stock exchange \cite{Gan2015}. A similar concept also applied in a movie recommendation system in the form of a dendrogram \cite{Vlachos2013}.

On the other side, Japanese animation, which is known as anime, has become internationally widespread nowadays. Not only in the eastern countries, but American audience also enjoying anime through Hayao Miyazaki's Studio Ghibli, which is well-known in western. 
According to Oricon's data \footnote{http://www.oricon.co.jp; data were gathered in HTTP://www.someanithing.com}, for the past five years (2011-2015), the Blu-ray Disc and DVD selling of anime were quite stable; it sold more than 600 thousand discs for each year. With the massive popularity of anime, a recommendation system is needed to find the similar anime based on particular indicators. 

This work aims to represent anime similarities using a minimum spanning tree to be used as a recommendation system. Moreover, the significance of anime will be revealed by extracting the centralities of the minimum spanning tree.

\section{Similarity Measurements}

A distance measurement between anime has to be calculated before constructing the minimum spanning tree. In this case, similarity measurement between anime will be used. Many works have been done in similarity measurements of movies in general. Researchers used user reviews \cite{Jacob2009}, movie mood \cite{Shi2012}, and movie score \cite{Takacs2007} to determine the similarity. There is also an NLP approach proposed using topic and summary similarity \cite{Fleischman2003}. In this work, besides using the score and topic, a new measurement is proposed, namely crew similarity. It is considered that crew similarity is an important characteristic in anime recommendation. Thus, the similarity measurements are described as follows:

\begin{enumerate}
	\item Crew Similarity
	
	There are two kinds of crew working in the anime industry, \textit{viz.}, production staff and voice actor/actress. Both of them are considered as important factors determining the anime success. Here a similarity measurement by using those factors is proposed, namely crew similarity. Let $S_n$ be a set of crew involved in anime $n^{th}$. The crew similarity between two anime is defined as the number of crew (both staff and voice actor/actress) who work for both anime, as calculated as follows: 
	\begin{eqnarray}\label{eq:crew}
	d_{ij} &&= \log | S_i \cap S_j | , \\ &&(i=0,1,2,...,k-2), \nonumber \\ &&(j=i+1,i+2,...,k-1) \nonumber
	\end{eqnarray}
	where $|S|$ and $k$ means the number of members in set $S$ and the total number of anime, respectively. Here, log transformation is applied since there are some data which are too far away from the others.

	\item Score Histogram Similarity
	
	Score histogram is determined by using user votes. There are some categories that user can select reflecting anime score. For instance, in Anime News Network, anime's votes are classified into 11 categories: Masterpiece, Excellent, Very good, Good, Decent, So-so, Not really good, Weak, Bad, Awful, and Worst ever. Based on the number of votes for each category, the total score of anime is calculated. Thus, here the votes for each group are assumed to be a histogram of scores. The similarity between score histogram would represent the user preference for the particular anime. Let $X_i$ be the score histogram of anime $i^{th}$ which is defined as
	\begin{equation}
	X_i = \{x_i^1, x_i^2, x_i^3, ..., x_i^N\},
	\end{equation}
	where
	\begin{equation}
	x_i^N = \frac{C_n}{\sum_{n=1}^{N} C_n},
	\end{equation}
	%	\begin{equation}
	%	X_i = \{ x_i^n | x_i^n = \frac{C_n}{\sum_{n} C_n}, \forall n \},
	%	\end{equation}
	and $C_n$ is the number of votes for category $n$, while N is the number of score categories. Then, the score histogram similarity ($s_{ij}$) is calculated using chi-squared distance as follows:
	\begin{equation}\label{eq:hist}
	s_{ij} = \sum_{x_i \in X_i, x_j \in X_j} \frac{(x_i - x_j)^2}{x_i + x_j}.
	\end{equation}

	\item Topic similarity
	
	Each anime is commonly labeled with some genres to make it easier to be classified. In Anime News Network, besides genre, anime are also categorized into themes. Both genre and theme are considered as critical parameters for classifying anime. Therefore, a topic similarity is used here, employing both genre and theme of anime, to show the similarity concerning the content. The topic similarity between two anime is defined as the number of topics (genres and themes) that present in both anime. Let $G_n$ be a set of genre and theme terms of anime $n^{th}$. Then topic similarity, $h_{ij}$, is calculated as
	\begin{equation} \label{eq:topic}
	h_{ij} = | G_i \cap G_j |
	\end{equation}
	
\end{enumerate}

Since the three measurements have different scales, a normalization size is performed. The calculation is given by
\begin{equation}
\hat{z}_{ij} = \frac{z_{ij} - z_{\text{min}}}{z_{\text{max}}-z_{\text{min}}}, (z = d,s,h) 
\end{equation}
where,
\begin{equation}
z_{\text{min}} = \underset{0\leq i,j \leq K-1}{\text{min}} z_{ij}, \hspace{5mm} z_{\text{max}} = \underset{0 \leq i,j \leq K-1}{\text{max}} z_{ij}, \hspace{5mm} (z = d,s,h)
\end{equation}
From Eq. (\ref{eq:crew}), (\ref{eq:hist}), and (\ref{eq:topic}) we know that crew and topic similarity result in higher values when both anime are considered similar, however, for the score histogram similarity, the result is otherwise. Hence, the crew and topic similarities are recalculated so that all similarities are aligned. Let $\hat{d}$, $\hat{s}$, and $\hat{h}$ be the normalized version of $d$, $s$, and $h$, respectively. The calculation is given by
\begin{equation}
\acute{q}_{ij} = 1 - \hat{q}_{ij}, (q = d,h)
\end{equation}

Afterwards, all three measurements are combined into a similarity vector, defined as $\mathbf{S}_{ij} = [\acute{d}_{ij},\hat{s}_{ij},\acute{h}_{ij}]'$, where $\mathbf{S}'$ means the transposition of $\mathbf{S}$. Thus total distance $\delta_{ij}$ is calculated as
\begin{equation}
\delta_{ij} = \| \mathbf{S}_{ij} \| 
\end{equation}
where $\| . \|$ is the Euclidean distance. Afterwards, in the next section, the total distance between anime, $\sigma_{ij}$, will be used as the edge length of a graph.

\section{Minimum Spanning Tree Representation}
In order to construct the minimum spanning tree, Kruskal's algorithm is employed (Alg. \ref{alg:kruskal}). The algorithm is implemented using disjoint-set data structure. Let $V = \{v_k, 0 \leq k \leq k-1\}$, be a set of vertices, $E = \{(v_i,v_j), 0 \leq i,j \leq k-1\}$ be a set of edges connecting a pair of vertices, and 
$w = \{\delta_{ij}, 0 \leq i,j \leq k-1 \}$ be a set of total distances obtained from the previous section.

\begin{algorithm}
	\caption{Kruskal's Algorithm}\label{alg:kruskal}
	\begin{algorithmic}[1]
		\Procedure{MakeSet}{$v$}
		\State Create new set containing $v$
		\EndProcedure \\
		\Function{FindSet}{$v$} 
		
		\Return a set containing $v$
		\EndFunction \\
		\Procedure{Union}{$u$,$v$}
		\State Unites the set that contain $u$ and $v$ into a new set
		\EndProcedure \\
		\Function{Kruskal}{$V,E,w$}
		\State $A \gets \{\}$
		\For {each vertex $v$ in $V$}
		\State MakeSet($v$)
		\EndFor
		\State Arrange $E$ in increasing costs, ordered by $w$
		\For {each ($u$,$v$) taken from the sorted list}
			\If {FindSet($u$) $\neq$ FindSet($v$)}
				\State $A \gets A \cup \{(u,v)\}$
				\State Union($u,v$)
			\EndIf
		\EndFor
		
		\Return A
		\EndFunction
	\end{algorithmic}
\end{algorithm}

In graph theory, measuring central vertex has been an active area of research. Many researchers proposed centrality indicators. Some of them are based on walk structure, namely degree \cite{Nieminen1974} and eigenvector centralities \cite{Bonacich1987}. Apart from it, there is also some which are based on geodesic distance, such as betweenness \cite{Freeman1977} and closeness centrality \cite{Freeman1978}. In this work, central vertices are going to be identified based on the indicators mentioned above. The centrality measurements are described below:
\begin{enumerate}
	
	\item Degree Centrality
	
	Degree centrality of vertex $v$ is the proportion of other vertices that are adjacent to $v$. It is defined as
	\begin{equation}
	C_D(v) = \frac{1}{k-1} \sum_{u \in V} a(u,v)
	\end{equation}
	where,
	\begin{equation}
	a(u,v) = \begin{cases}
	& 1, \text{if $u$ and $v$ are connected by a line} \\
	& 0, \text{otherwise}
	\end{cases}
	\end{equation}
	Anime having high degree means that it have many similar anime around it. 
	
	\item Eigenvector Centrality
	
	Conceptually, eigenvector centrality is similar to degree centrality. The centrality also measures the number of walks of a vertex. However, instead of having the length of one, eigenvector measures the number of walks of length infinity. Thus, in eigenvector centrality, a vertex has a high centrality if it is connected to another vertex that also has a high centrality. The eigenvector centrality is defined as the summed connection of a vertex to others, weighted by their centralities.  
	Let $R = {r_{uv}}$ be a matrix of relationship, \textit{i.e.}, $r_{uv} = a(u,v)$. Eigenvector centrality of vertex $v$, denoted as $e_v$, is calculated as follows
	\begin{equation}
	\lambda e_v = \sum_u r_{uv}e_u
	\end{equation}
	where $\lambda$ is a constant required so that the equation have a non zero solution. This problem can be rewritten as an eigenvector equation:
	\begin{equation}
	\lambda \textbf{e} =  R \textbf{e}
	\end{equation}
	where $\textbf{e}$ is the eigenvector of $R$ and $\lambda$ is the corresponding eigenvalue.
	
	\item Betweenness Centrality
	
	In the concept of betweenness centrality, a vertex is called in the central position when it is located on the shortest path between two vertices. Based on it, betweenness centrality of a vertex is defined as the number of times a vertex acts as a bridge between the shortest path of two other vertices. The betweenness centrality of vertex $v$ is defined as
	
	\begin{equation}
	C_B(v) = \sum_{s \neq v \neq t \in V} \frac{\sigma_{st}(v)}{\sigma_{st}}
	\end{equation}
	where $\sigma_{st}$ is the number of shortest paths between $s$ and $t$, $\sigma_{st}(v)$ is the fraction of those shortest paths that pass through $v$. According to Freeman \cite{Freeman1977}, the betweenness can be normalized as
	\begin{equation}
	\hat{C}_B(v) = \frac{2C_B(v)}{k^2 - 3k + 2}
	\end{equation}
	
	\item Closeness Centrality
	
	Closeness measures how close a vertex to all other vertices in a graph. It is defined as the inverse of the total distance from a vertex to other vertices. Since this measurement depends on the number of vertices in the graph, the relative closeness is calculated with a normalization. The relative closeness is given by
	
	\begin{equation}
	C_C(v) = \frac{k-1}{\sum_{u = 1}^{k-1} d(v,u)}
	\end{equation}
	where $d(v,u)$ is the shortest-path distance between $v$ and $u$. High value of closeness means that a vertex is relatively close to the other vertices.  
	
\end{enumerate}
In order to calculate the total centrality, Euclidean distance are applied so that $\varphi (v) = \sqrt{C_D(v)^2 + e_v^2 + \hat{C}_B(v)^2 + C_c(v)^2}$, where $\varphi (v)$ is the total centrality of vertex $v$.

\section{Results}
\begin{figure}[t]
	\centering
	\includegraphics[width=2in]{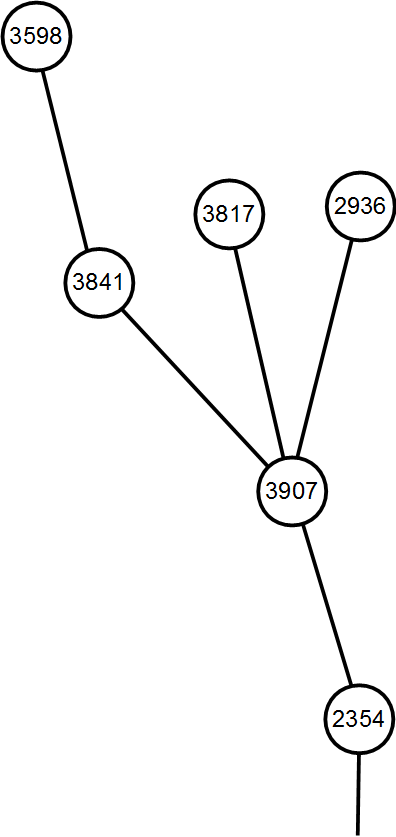}
	\caption{Sample of a small branch of the minimum spanning tree.}
	\label{fig:smallbranch}
\end{figure}

\begin{figure*}[!ht]
	\centering
	\includegraphics[width=5in]{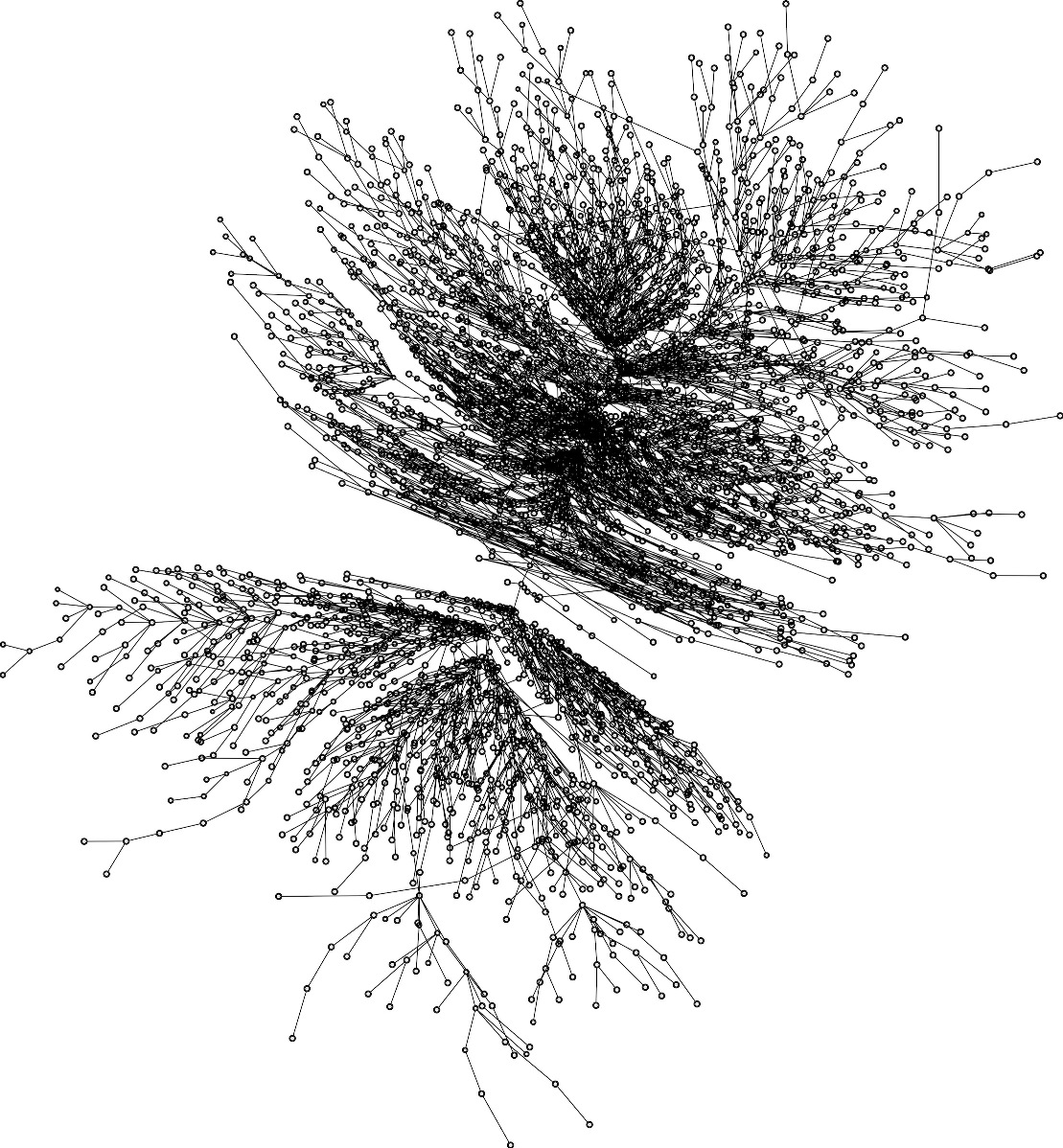}
	\caption{The minimum spanning tree result.}
	\label{fig:mst}
\end{figure*}

In this work, 4029 anime data were collected randomly from Anime News Network\footnote{http://www.animenewsnetwork.com, accessed May 10, 2016}. For each anime pair, three similarity measurements are calculated, then used as features to build the minimum spanning tree representation. To visualize the minimum spanning tree, a neato program from graphviz is employed \cite{North2004}. Neato program constructs a spring layout by minimizing the global energy function \cite{Kamada1989}. The resulting minimum spanning tree is shown in Fig. \ref{fig:mst}. The sample of a branch of the minimum spanning tree is shown in Fig. \ref{fig:smallbranch}. It can be seen that using the minimum spanning tree; we can obtain the nearest anime. For instance, anime number 3907 has four similar anime, namely anime number 3841, 3817, 2936 and 2354. This information is useful for recommender systems. Furthermore, the farness between anime also can be retrieved. If we want to know the distance between anime number 3598 and 2354, it can be seen from Fig. \ref{fig:smallbranch}, that the distance is three walks through anime number 3841 and 3907. It is considered as a small distance if we take a look at the overall minimum spanning tree in Fig. \ref{fig:mst}.

\begin{figure}[t]
	\centering
	\subfigure[Degree Distribution]{\includegraphics[width=1.7in]{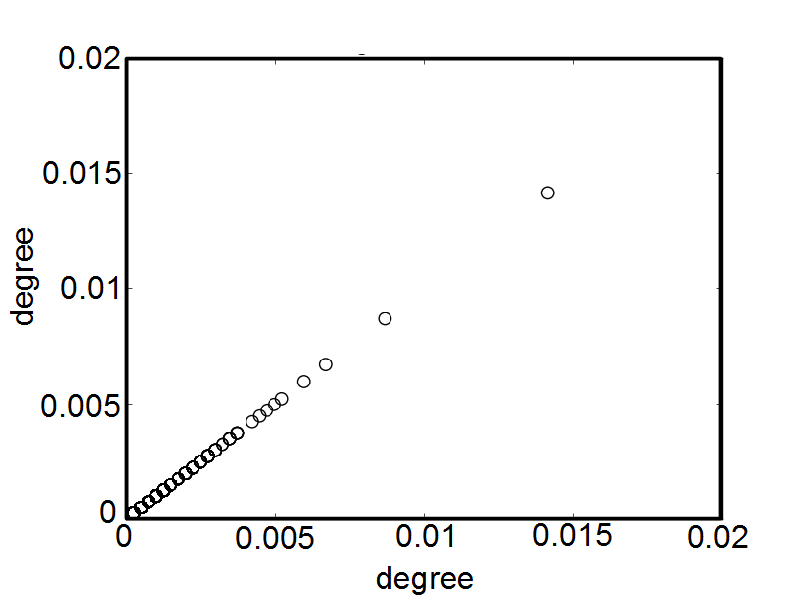}}
	\subfigure[Eigenvector Distribution]{\includegraphics[width=1.7in]{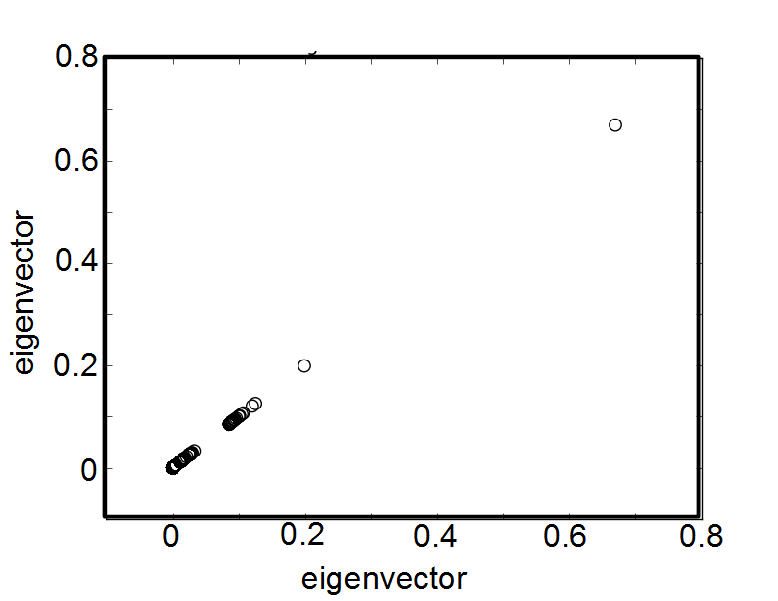}}
	\subfigure[Betweenness Distribution]{\includegraphics[width=1.7in]{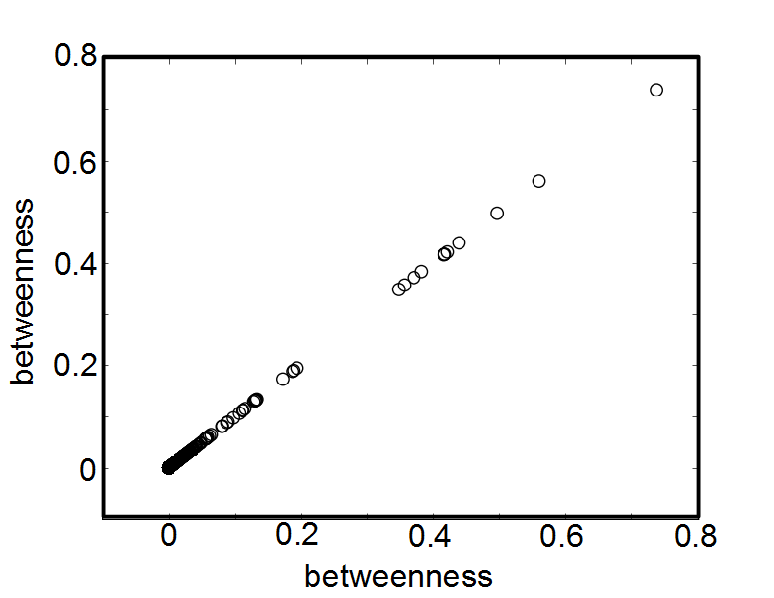}}
	\subfigure[Closeness Distribution]{\includegraphics[width=1.7in]{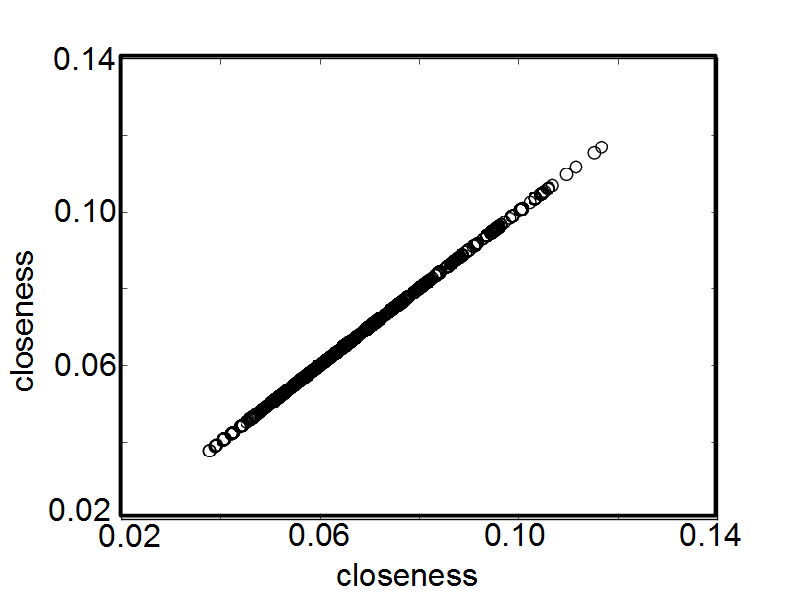}}
	\caption{Distribution for each centrality measurements}
	\label{fig:centrality}
\end{figure}

Distribution of each centrality measurement is shown in Fig. \ref{fig:centrality}. The degree distribution is shown in Fig. \ref{fig:centrality}(a). It can be seen that most points are located at the bottom-left corner. This means the majority of vertices are having small numbers of degree. In other words, only a small number of anime that have numerous similar anime surround it. 
In case of eigenvector distribution, shown in Fig. \ref{fig:centrality}(b), only one value which has centrality and separated quite far from others. This means that the minimum spanning tree tends towards one direction. The corresponding anime has such a great importance in the graph. 
The distribution of betweenness is shown in Fig. \ref{fig:centrality}(c). It can be seen that some points have large betweenness, but many points are otherwise. 
Closeness centrality distribution is shown in Fig. \ref{fig:centrality}(d). The points are distributed from the small one until the largest closeness. This kind of distribution is expected for closeness. Vertices that are located in the outer part of the graph are having the small value of closeness. Nevertheless, large values are obtained as the vertices located near the center of the graph. Thus, many of the vertices are located between the outer and the center of the graph.

Table \ref{tab:centrality} shows the corresponding anime having the most significant centrality value for each measurement. It can be seen that anime One Piece and Naruto are ranked the first and second respectively in all centrality measurements as well as the whole one. It shows the significance of those anime among others. 

\begin{table*}[!ht]
	\centering
	\caption{Highest values of each centrality}
	\label{tab:centrality}
	\begin{tabular}{clllll}
		\hline \hline
		Rank & Degree & Eigenvector & Betweenness & Closeness  & Total                                      \\
		\hline \hline
		1    & One Piece & One Piece & One Piece & One Piece & One Piece \\
		2    & Naruto  & Naruto & Naruto & Naruto & Naruto     \\
		3    & Aquarion Evol & Detective Conan & Bleach & Bleach & Bleach \\    
		\hline 
	\end{tabular}
\end{table*}

\section{Concluding Remarks}
A new way of representing anime similarity has been proposed by applying the minimum spanning tree. Here, we used similarity measurement, called crew similarity, as an addition to the commonly used similarity measurements, namely scores and topics. The results show that using a minimum spanning tree; a similar anime can be obtained easily by looking at the graph. Moreover, we found that anime such as One Piece, Naruto, and Bleach are considered as the essential anime based on the centrality calculations.

%\bibliographystyle{srt}
%\bibliographystyle{ieeetr}
%\bibliography{} 

\begin{thebibliography}{10}
	
\bibitem{Herman2000}
I.~Herman, G.~Melancon, and M.~S. Marshall, ``Graph visualization and
navigation in information visualization: A survey,'' {\em IEEE Transactions
	on Visualization and Computer Graphics}, vol.~6, pp.~24--43, Jan 2000.

\bibitem{Moncla2016}
L.~Moncla, M.~Gaio, J.~Nogueras-Iso, and S.~Mustière, ``Reconstruction of
itineraries from annotated text with an informed spanning tree algorithm,''
{\em International Journal of Geographical Information Science}, vol.~30,
no.~6, pp.~1137--1160, 2016.

\bibitem{Murmu2015}
M.~K. Murmu, ``A distributed approach to construct minimum spanning tree in
cognitive radio networks,'' {\em Procedia Computer Science}, vol.~70, pp.~166
-- 173, 2015.
\newblock Proceedings of the 4th International Conference on Eco-friendly
Computing and Communication Systems.

\bibitem{Crobe2016}
A.~Crobe, M.~Demuru, L.~Didaci, G.~L. Marcialis, and M.~Fraschini, ``Minimum
spanning tree and k -core decomposition as measure of subject-specific eeg
traits,'' {\em Biomedical Physics and Engineering Express}, vol.~2, no.~1,
p.~017001, 2016.

\bibitem{Maican2015}
V.~MAICAN, ``Minimum spanning tree algorithm on mapreduce one-chip
architecture,'' {\em Romanian Journal of Information Science and Technology},
vol.~18, no.~2, pp.~126--143, 2015.

\bibitem{Gan2015}
S.~L. Gan and M.~A. Djauhari, ``New york stock exchange performance: evidence
from the forest of multidimensional minimum spanning trees,'' {\em Journal of
	Statistical Mechanics: Theory and Experiment}, vol.~2015, no.~12, p.~P12005,
2015.

\bibitem{Vlachos2013}
M.~Vlachos and D.~Svonava, ``Recommendation and visualization of similar movies
using minimum spanning dendrograms,'' {\em Information Visualization},
vol.~12, no.~1, pp.~85--101, 2013.

\bibitem{Jacob2009}
N.~Jakob, S.~H. Weber, M.-C. Müller, and I.~Gurevych, ``Beyond the stars:
Exploiting free-text user reviews to improve the accuracy of movie
recommendations,'' in {\em Proceedings of the 1st international CIKM workshop
	on Topic-sentiment analysis for mass opinion}, pp.~57--64, 2009.

\bibitem{Shi2012}
Y.~Shi, M.~Larson, and A.~Hanjalic, ``Mining mood-specific movie similarity
with matrix factorization for context-aware recommendation,'' in {\em
	Proceedings of the workshop on context-aware movie recommendation},
pp.~34--40, ACM, 2010.

\bibitem{Takacs2007}
G.~Takacs, I.~Pilaszy, B.~Nemeth, and D.~Tikk, ``On the gravity recommendation
system,'' in {\em Proceedings of KDD Cup Workshop at SIGKDD’07, 13th ACM
	Int. Conf. on Knowledge Discovery and Data Mining}, pp.~22--30, 2007.

\bibitem{Fleischman2003}
M.~Fleischman and E.~Hovy, ``Recommendations without user preferences: a
natural language processing approach,'' in {\em Proceedings of the 8th
	international conference on Intelligent user interfaces}, pp.~242--244, ACM,
2003.

\bibitem{Nieminen1974}
J.~Nieminen, ``On the centrality in a graph,'' {\em Scandinavian journal of
	psychology}, vol.~15, no.~1, pp.~332--336, 1974.

\bibitem{Bonacich1987}
P.~Bonacich, ``Power and centrality: A family of measures,'' {\em American
	journal of sociology}, pp.~1170--1182, 1987.

\bibitem{Freeman1977}
L.~C. Freeman, ``A set of measures of centrality based on betweenness,'' {\em
	Sociometry}, pp.~35--41, 1977.

\bibitem{Freeman1978}
L.~C. Freeman, ``Centrality in social networks conceptual clarification,'' {\em
	Social networks}, vol.~1, no.~3, pp.~215--239, 1978.

\bibitem{North2004}
S.~C. North, {\em Drawing graphs with NEATO}, http://www.graphviz.org/pdf/neatoguide.pdf, 2004

\bibitem{Kamada1989}
T.~Kamada and S.~Kawai, ``An algorithm for drawing general undirected graph,''
{\em Information Processing Letters}, vol.~31, pp.~7--15, 1989.

	
\end{thebibliography}

\end{document}